\documentclass[reprint,superscriptaddress,nofootinbib,prd,floatfix]{revtex4}
\pdfoutput=1
\usepackage{graphicx}
\usepackage[dvipsnames]{xcolor}
\usepackage{amsmath}
\allowdisplaybreaks
\usepackage{amssymb}

\usepackage{slashed}
\usepackage[utf8]{inputenc}
\usepackage[colorlinks=true,linkcolor=blue,bookmarksopen,bookmarksnumbered]{hyperref}
\usepackage{color} % 它已经被 dvipsnames 选项定义了

\usepackage[normalem]{ulem}
\usepackage{soul}
\usepackage{units}
\usepackage{rotating}
\usepackage{hhline,multirow,tabularx}
\usepackage{bm}

\usepackage{cleveref} % cleveref 在 nameref 之前加载
\usepackage{nameref} % nameref 最好放在 hyperref 之后
\usepackage[export]{adjustbox}
\usepackage{mdframed}
\usepackage{comment}
\usepackage{natbib}
\usepackage{aliascnt} % aliascnt 放在最后

\def\bea#1\eea{\begin{align}#1\end{align}}

\usepackage{lipsum}
\usepackage{hyperref}
\begin{document}

\title{Derivation of the Balitsky-Kovchegov Equation for Quark-Quark Scattering}

\author{Cong Li}
\affiliation{School of Information Engineering, Zhejiang Ocean University, Zhoushan, 316022, Zhejiang, China}

%%%%%%%%%%%%%%%%%%%%%%%%%%%%%%%%%%%%%%%%%%%%%%%%%%%%%%%%%%%%%%

\begin{abstract}

We derived the BK equation for quark-quark scattering, extending the dipole-hadron scattering framework. The new BK equation reveals that the quark-quark scattering amplitude increases with increasing quark rapidity. Since the momentum-dot product is Lorentz-invariant, the coupling constant plays a crucial role in accounting for it. The new BK equation may describe how the coupling constant depends on the rapidity.

\end{abstract}
\maketitle

\section{Introduction}

In quantum chromodynamics (QCD), higher-order corrections from virtual particle loops are crucial to understanding the running of the coupling constant \cite{schwartz}. These loop corrections alter the coupling constant, leading to renormalization. The renormalization group equation describes how the coupling constant varies with energy scale, known as "running". This running has significant implications: it not only helps verify QCD's predictions in high-energy physics experiments, but also aids in understanding conditions in the early universe. By studying these higher-order corrections, we can make more accurate experimental predictions and deepen our understanding of elementary particles and their interactions.

The Balitsky-Kovchegov (BK) equation describes the scattering of a dipole (quark-antiquark pair) with hadrons as a function of the dipole's rapidity \cite{Bk1,BK2,BK3}. It accounts for nonlinear effects of the high-density gluon field in hadrons at high energies, modeling gluon recombination to describe how the dipole's scattering amplitude evolves with changes in energy (or momentum fraction $ x $ ). By combining Balitsky's recursive functional relation with Kovchegov's simplifications, the equation becomes numerically tractable and effectively illustrates the evolution of the gluon distribution toward saturation at increasing energies.

In deriving the BK equation, we focus on the dynamics of scattering between a dipole and a hadron \cite{BKd1,Bkd2}. First, the virtual photon emitted by the electron generates the dipole. For the scattering amplitude of the dipole and hadron at rapidity $y + \Delta y$, there are two distinct approaches to obtain it. The first approach considers direct scattering of the dipole and the hadron, resulting in a scattering amplitude that depends on $y + \Delta y$. In the second approach, a gluon is emitted after boosting the dipole to rapidity $y$. In this case, the system consisting of the dipole and the gluon scatters with the hadron, and the scattering amplitude is now a function of $y$ rather than $y + \Delta y$. Since the physical process remains invariant regardless of the chosen approach, we obtain the BK equation, which describes the evolution of the scattering amplitude with respect to rapidity. It is important to note that the scattering probability of a dipole of a given size in the target field is directly determined by the gluon distribution of the target. Thus, the evolution of the scattering amplitude with rapidity can generally be interpreted as the evolution of the gluon distribution with respect to rapidity.

Based on the scattering dynamics between a dipole and a hadron \cite{BKd1,Bkd2}, which underlies the derivation of the BK equation, we propose to derive the BK equation specifically for quark-quark scattering. When a projectile quark collides with a target quark, there are also two distinct approaches to obtain it. Firstly, the projectile quark directly scatters with the target quark at rapidity $y + \Delta y$. Secondly, the projectile quark emits a gluon due to a boost, and subsequently the quark-gluon system scatters with the target quark at rapidity $y$. Given the independence of the scattering amplitude from the chosen approach, we ultimately derive the evolution equation for the amplitude with respect to rapidity in the context of quark-quark scattering, which we term the BK equation for quark-quark scattering. 

The structure of this paper is very simple. In the next section, we derive the BK equation for the quark-quark scattering based on the BK equation for the dipole-hadron scattering. In the end, a brief summary will be given.

\section{BK equation for quark-quark scattering}

%%\label{}

Firstly, we derive the probability that a quark emits a gluon, as shown in Figure 1.
 \begin{figure}[ht]
	\centering 
	\includegraphics[width=0.3\textwidth, angle=0]{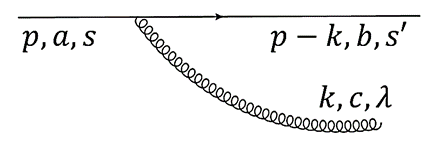}	
	\caption{The initial quark emits a gluon.}
 %\label{11}%
\end{figure}
where the initial quark with momentum $ p $, spin $ s $, and color $ a $, emits a gluon with momentum $ k $, color $ c $, and helicity $ \lambda $. Then, the momentum, spin, and color of the final quark are $ p - k $, $ s' $, and $ b $, respectively. In the high-energy limit, the $p^+$ of the initial quark light-cone momentum is larger, and the emitted gluon is very soft. Using Feynman's rule, the amplitude of this process is \cite{Light-cone},
\begin{equation}
\psi_{q \rightarrow q g}\left(k^{+}, k_T\right)=\frac{\bar{u}_{s^{\prime}}(p-k)}{\sqrt{(2 \pi)^3 2(p-k)^{+}}} \frac{t_{a b}^c g \varepsilon_\lambda^\mu(k) \gamma_\mu}{\sqrt{(2 \pi)^3 2 k^{+}}} \frac{u_s(p)}{\sqrt{(2 \pi)^3 2 p^{+}}} \frac{(2 \pi)^3}{p^{-}-k^{-}-(p-k)^{-}}
\end{equation}
where $(2\pi)^3 (p_{\text{initial}}^{-} - p_{\text{final}}^{-})^{-1} = (2\pi)^3 (p^{-} - k^{-} - (p-k)^{-})^{-1}$ is the light-cone energy denominator, and $(p-k)^{-}$ represents the minus component of the light-cone momentum of the on-shell particle with 3-momentum $\vec{p} - \vec{k}$. Using the conditions that quarks and gluons are on-shell and that gluon is soft in the high-energy limit, the amplitude can be simplified to.
\begin{equation}
\psi_{q\rightarrow q\ g}\left(k^+,k_T\right)=-\frac{\sqrt2g}{\sqrt{(2\pi)^3}}t_{ab}^c\frac{1}{\sqrt{k^+}}\frac{k_T\cdot\varepsilon_T}{k_T^2}\delta_{s,s^\prime}
\end{equation}
After Fourier transformation, we get the amplitude in the transverse-coordinate space  \cite{Light-cone}.
\begin{equation}
\psi_{q\rightarrow q\ g}\left(k^+,r_T\right)=\int\frac{d^2k_T}{\sqrt{(2\pi)^2}}e^{ik_T\cdot r_T}\psi_{q\rightarrow q\ g}\left(k^+,k_T\right)\ =-i\frac{\sqrt2g}{\sqrt{(2\pi)^3}}t_{ab}^c\frac{1}{\sqrt{k^+}}\frac{r_T\cdot\varepsilon_T}{r_T^2}\delta_{s,s^\prime}
\end{equation}
  Secondly, we can derive the wavefunction of a single quark. First, we can now write the Fock state wavefunction of the quark as
\begin{equation}
\left|q_a\right\rangle=C\left|q_a\left(x\right)\right\rangle_0+\int dk^+d^2r_T\psi_{q\rightarrow qg}\left(k^+,r_T\right)\left|q_b(x)g_c(y)\right\rangle_0
\end{equation}
The wavefunction of the quark is set to be normalized, with $C$ used to normalize the wavefunction of the quark at the next leading order. Its inner product is given by.
\begin{equation}
\left\langle q_a\middle|\ q_a\right\rangle=\left|C\right|^2+\int\ dk^+d^2r_T\frac{g^2}{4\pi^3k^+}t_{\alpha b}^ct_{b\alpha}^c\left|\varepsilon_T^\lambda.\frac{r_T}{r_T^2}\right|^2=\left|C\right|^2+\int\ d^2r_T\frac{dz}{z}\frac{g^2}{4\pi^3}\frac{N_c^2-1}{2}\frac{1}{r_T^2}
\end{equation}
In the last identity, we use the identities,
\begin{equation}
    \sum_{\lambda=1,2}{(\varepsilon_\lambda^\ast\cdot x)(\varepsilon_\lambda\cdot y)} = x\cdot y,
\end{equation}
\begin{equation}
    t_{ab}^c t_{ba}^c = \frac{N_c^2-1}{2},
\end{equation}
\begin{equation}
    k^+ = z\ P^+.
\end{equation}
where $z$ denotes the momentum fraction carried by the gluon. As $\left\langle q_a\middle|\ q_a\right\rangle=1$, we obtain,
\begin{equation}
\left|C\right|^2=1-\frac{\alpha_s{(N}_c^2-1)}{2\pi^2}\int\ d^2r_Tdy\frac{1}{r_T^2}
\end{equation}
where the relationship between rapidity and momentum fraction is $y=\ln{(1/z)}$ and the coupling constant $\alpha_s=g^2/(4\pi)$.
\begin{figure}[ht]
	\centering 
	\includegraphics[width=0.7\textwidth, angle=0]{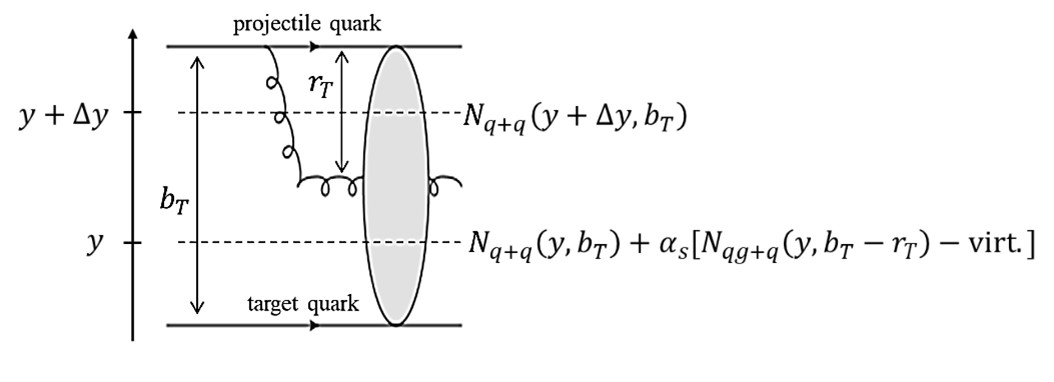}	
	\caption{The emitted gluon can be seen as a part of the target quark wavefunction or as a part of the projectile quark wavefunction.}
 %\label{11}%
\end{figure}

Thirdly, considering all the aforementioned points, we shall proceed to the main topic and derive the BK equation for quark-quark scattering. This methodology has previously been employed in deriving the BK equation for dipole-hardon scattering. First, for the amplitude between a projectile quark and a target quark, we boost the projectile quark from rapidity $y$ to $y + \Delta y$. This boosting process creates a phase space region where a gluon can be emitted. The emitted gluon can be regarded as a component of the projectile quark's wavefunction. In this way, the scattering amplitude of the projectile quark off the target quark is.
\begin{equation}
\begin{aligned}
    &\left|C\right|^2N_{q+q}\left(y,b_T\right)+\frac{\alpha_s(N_c^2-1)}{2\pi^2}\int dy d^2r_T\frac{1}{r_T^2}N_{qg+q}\left(y,r_T,b_T\right)\ \\
    =&N_{q+q}\left(y,b_T\right)+\frac{\alpha_s{(N}_c^2-1)}{2\pi^2}\int dy d^2 r_T\frac{1}{r_T^2}\left[N_{qg+q}\left(y,\ r_T,b_T\ \right)-N_{q+q}\left(y,b_T\ \right)\right]
\end{aligned}
\end{equation}
where $N_{q+q}\left(y,b_T\right)$ denotes the amplitude for the projectile quark-target quark scattering, and $b_T$ is the impact parameter. Similarly, $N_{qg+q}\left(y,b_T\right)$ denotes the amplitude for the system consisting of the projectile quark and a gluon against the target quark. $r_T$ refers to the size of the quark-gluon system.

The amplitude discussed previously corresponds to the lower dashed line depicted in Figure 2. Alternatively, the same process can be interpreted in a different way. Specifically, we can consider the gluon as an integral part of the target quark wavefunction. Consequently, the process simplifies to the scattering of two quarks at a rapidity of $y + \Delta y$, with the corresponding scattering amplitude denoted as $N_{q+q}\left(y+\Delta y,b_T\right)$. This interpretation aligns with the upper dashed line in Figure 2. Given that physical observables are independent of the chosen interpretation, both descriptions are the same underlying physical process. On this basis, we derive a renormalization group equation.
\begin{equation}
N_{q+q}\left(y+\Delta y,b_T\right)=N_{q+q}\left(y,b_T\right)+\frac{\alpha_s{(N}_c^2-1)}{2\pi^2}\Delta y\int{d^2r_T}\frac{1}{r_T^2}\left[N_{qg+q}\left(y,\ r_T,b_T\ \right)-N_{q+q}\left(y,b_T\ \right)\right]
\end{equation}
By subtracting $N_{q+q}\left(y,b_T\right)$ from both sides of the equation, dividing the result by $ \Delta y$, and then taking the limit as $\Delta y \rightarrow 0$, we obtain.
\begin{equation}
\partial_y N_{q+q}\left(y,b_T\right)=\frac{\alpha_s{(N}_c^2-1)}{2\pi^2}\int{d^2r_T}\frac{1}{r_T^2}\left[N_{qg+q}\left(y,\ r_T,b_T\ \right)-N_{q+q}\left(y,b_T\ \right)\right]
\end{equation}
This equation reveals that the quark-quark scattering amplitude is not Lorentz invariant. It comprises a real contribution originating from gluon radiation and a virtual contribution $-N_{q+q}\left(y,b_T\ \right)$, which arises due to the normalization requirement of the wavefunction. Since it is derived from the BK equation for dipole-hadron scattering, we refer to it as the BK equation for quark-quark scattering. It implicitly contains the evolution of the strong coupling constant with rapidity, and the proof is as follows. First, the Fourier Transform to the equation 12 is,
\begin{equation}
\int\frac{d^2b_T}{2\pi}\ e^{iq_T\cdot b_T}\ \partial_yN_{q+q}\left(y,b_T\right)=\frac{\alpha_s{(N}_c^2-1)}{2\pi^2}\int\frac{d^2b_T}{2\pi}\ e^{iq_T\cdot b_T}\int{d^2r_T}\frac{1}{r_T^2}\left[N_{qg+q}\left(y,\ r_T,b_T\ \right)-N_{q+q}\left(y,b_T\ \right)\right]
\end{equation}
we obtain the scattering amplitude in momentum space then.
\begin{equation}
\partial_y N_{q+q} (y,q_T) = \frac{\alpha_s (N_c^2-1)}{2\pi^2} \int d^2 r_T \frac{1}{r_T^2} \left[N_{qg+q} (y, r_T,q_T) - N_{q+q} (y,q_T)\right]
\end{equation}
It's also easy to get the conjugate amplitude's evolution equation.
\begin{equation}
\partial_y N_{q+q}^* (y,q_T) = \frac{\alpha_s (N_c^2-1)}{2\pi^2} \int d^2 r_T \frac{1}{r_T^2} \left[N_{qg+q}^* (y, r_T,q_T) - N_{q+q}^* (y,q_T)\right]
\end{equation}
When discussing the evolution equation of the scattering cross-section with rapidity, we have.
\begin{equation}
\partial_y \left[N_{q+q}^* (y,q_T) N_{q+q} (y,q_T)\right] = N_{q+q}^* (y,q_T) \partial_y N_{q+q} (y,q_T) + N_{q+q} (y,q_T) \partial_y N_{q+q}^* (y,q_T)
\end{equation}
Next, for the evolution equation of the scattering cross-section with rapidity, there is an another interpretation. For the scattering amplitude $N_{q+q} (y,q_T) $, its leading-order contribution is $N_{q+q} (y,q_T) = ig_s^2 \left[\overline{u}(p_3) \gamma^\mu T^a u(p_1)\right] \frac{g_{\mu\nu}}{q^2} \left[\overline{u}(p_4) \gamma^\nu T^a u(p_2)\right]$. The coupling constant in the leading-order amplitude $N_{q+q}(y,q_T) $  can be factored out,
\begin{equation}
N_{q+q} (y,q_T) = \alpha_s \mathcal{N}(s,t,u)
\end{equation}
where $\mathcal{N}(s,t,u)$ is only the function of the Mandelstam variables. Reexamining the evolution equation of the scattering cross-section through Equation 17, we have.
\begin{equation}
\partial_y \left[N_{q+q}^* (y,q_T) N_{q+q} (y,q_T)\right] = \partial_y \left[\alpha_s^2 |\mathcal{N}(s,t,u)|^2\right] = |\mathcal{N}(s,t,u)|^2 \partial_y \alpha_s^2
\end{equation}
In \(\partial_y \left[\alpha_s^2 |\mathcal{N}(s,t,u)|^2\right] = |\mathcal{N}(s,t,u)|^2 \partial_y \alpha_s^2\), we neglect \(\partial_y |\mathcal{N}(s,t,u)|^2\) because \(s, t, u\) are Lorentz scalars. Finally, combining Equation 16 with Equation 18, we obtain.
\begin{equation}
|\mathcal{N}(s,t,u)|^2 \partial_y \alpha_s^2 = N_{q+q}^* (y,q_T) \partial_y N_{q+q} (y,q_T) + N_{q+q} (y,q_T) \partial_y N_{q+q}^* (y,q_T)
\end{equation}
Or in a more concise form, it is.
\begin{equation}
\partial_y \alpha_s = \frac{\alpha_s}{2|N_{q+q} (y,q_T)|^2} \left[N_{q+q}^* (y,q_T) \partial_y N_{q+q} (y,q_T) + N_{q+q} (y,q_T) \partial_y N_{q+q}^* (y,q_T)\right]
\end{equation}
Evidently, this equation describes the evolution of the strong coupling constant with rapidity. It's important to note that Equation 19 holds under the assumption that all \( N_{q+q} (y,q_T) \) only consider the leading-order contributions. This phenomenon does not manifest in Quantum Electrodynamics (QED) processes because photons are uncharged, rendering the electron-photon system indistinguishable from a single electron. Further processing of this equation is challenging, not only because it is an integro-differential equation, but also because it contains $N_{qg+q} (y, r_T, q_T) $ which is the scattering amplitude of the quark-gluon system.

Alternatively, we can utilize the description of the BK equation to comprehend the process, too. The amplitude $N$ typically represents the distribution function of the gluon surrounding the target quark. As the projectile quark traverses the gluon field, quarks of varying rapidities encounter differing gluon densities, ultimately resulting in distinct cross sections. This effect can be incorporated into the coupling constant, where the strong-interaction coupling constant increases with increasing rapidity. In color glass condensate (CGC) theory, a saturation scale exists to constrain the potentially infinite increase. Similarly, the strong-interaction coupling constant may not escalate indefinitely with growing rapidity.

Finally, we deal with $N_{qg+q}\left(y,\ r_T,b_T\ \right)$. Under the large-$N_c$ approximation, the number of possible color states of the gluon is $N_c^2 - 1 \approx N_c^2$. Therefore, the gluon can be considered equivalent to a quark-antiquark pair \cite{forshaw1997quantum}, since the color state of a single quark is $N_c$. Based on this idea, the system $qg$ can be considered as $qq \bar{q}$. Consequently, the probability that the system does not scatter off the target quark is,
\begin{equation}
S_{qg+q}\left(b_T,r_T\right)=S_{q+q}\left(b_T\right)S_{q+q}\left(b_T-r_T\right)S_{\bar{q}+q}\left(b_T-r_T\right)
\end{equation}
where $S = 1-N$ is a probability not to scatter, we then have,
\begin{equation}
\begin{aligned}
    N_{qg+q}\left(b_T,r_T\right)&=N_{q+q}\left(b_T\right)+N_{q+q}\left(b_T-r_T\right)+N_{\bar{q}+q}\left(b_T-r_T\right)-N_{q+q}\left(b_T\right)N_{q+q}\left(b_T-r_T\right)\\
    &-N_{q+q}\left(b_T-r_T\right)N_{\bar{q}+q}\left(b_T-r_T\right)-N_{q+q}\left(b_T\right)N_{\bar{q}+q}\left(b_T-r_T\right)
\end{aligned}
\end{equation}
where we drop the higher-order terms $N_{q+q}\left(b_T\right)N_{q+q}\left(b_T-r_T\right)N_{\bar{q}+q}\left(b_T-r_T\right)$, because in perturbation theory, its contribution is negligible. Finally, we get the simpler BK equation.
\begin{equation}
\begin{aligned}
\partial_y N_{q+q}\left(y,b_T\right)=\frac{\alpha_sN_c^2}{2\pi^2}\int{d^2r_T}\frac{1}{r_T^2}&\left[N_{q+q}\left(y,b_T-r_T\right)+N_{\bar{q}+q}\left({y,b}_T-r_T\right)-N_{q+q}\left(y,b_T\right)N_{q+q}\left(y,b_T-r_T\right)\right.\\
&\left.-N_{q+q}\left(y,b_T-r_T\right)N_{\bar{q}+q}\left(y,b_T-r_T\right)-N_{q+q}\left(y,b_T\right)N_{\bar{q}+q}\left(y,b_T-r_T\right)\right]
\end{aligned}
\end{equation}
In this equation, $N_{q+q}\left(b_T\right)$ is eliminated by the virtual contribution term, and the remaining terms come from the scattering between the emitted gluons and the target quark.

\section{Summary}
%%\label{}

This paper derives the BK equation for quark-quark scattering, building upon the existing BK equation for dipole-hadron scattering. This equation elucidates the evolution of the quark-quark scattering amplitude with respect to rapidity, revealing that the quark-quark scattering amplitude is not Lorentz-invariant. Based on the quark-quark scattering BK equation, we further derived a equation for the evolution of the strong coupling constant with respect to rapidity. The equation not only demonstrates the running of the strong coupling constant with rapidity but also reveals the possibility of incorporating high-energy evolution effects into the coupling constant itself. In the Introduction, we described how high-order loop diagrams cause the coupling constant evolving with energy scale. Here, high-order Feynman diagrams that the radiation of real gluons lead to the coupling constant evolving with rapidity. Specifically, since gluons carry color charge, the scattering of the quark-gluon system with the target quark introduces a correction to the quark-quark scattering. This correction is absent in QED because photons are chargeless.

\section*{Acknowledgments}
AI tools were used to improve sentence fluency and make English more natural.

\bibliography{ref}

\begin{thebibliography}{8}
\expandafter\ifx\csname natexlab\endcsname\relax\def\natexlab#1{#1}\fi
\expandafter\ifx\csname bibnamefont\endcsname\relax
  \def\bibnamefont#1{#1}\fi
\expandafter\ifx\csname bibfnamefont\endcsname\relax
  \def\bibfnamefont#1{#1}\fi
\expandafter\ifx\csname citenamefont\endcsname\relax
  \def\citenamefont#1{#1}\fi
\expandafter\ifx\csname url\endcsname\relax
  \def\url#1{\texttt{#1}}\fi
\expandafter\ifx\csname urlprefix\endcsname\relax\def\urlprefix{URL }\fi
\providecommand{\bibinfo}[2]{#2}
\providecommand{\eprint}[2][]{\url{#2}}

\bibitem[{\citenamefont{Schwartz}(2013)}]{schwartz}
\bibinfo{author}{\bibfnamefont{M.~D.} \bibnamefont{Schwartz}}, \emph{\bibinfo{title}{Quantum Field Theory and the Standard Model}} (\bibinfo{publisher}{Cambridge University Press}, \bibinfo{address}{Cambridge, UK}, \bibinfo{year}{2013}), ISBN \bibinfo{isbn}{9781107034730}.

\bibitem[{\citenamefont{Balitsky}(1996)}]{Bk1}
\bibinfo{author}{\bibfnamefont{I.}~\bibnamefont{Balitsky}}, \bibinfo{journal}{Nuclear Physics B} \textbf{\bibinfo{volume}{463}}, \bibinfo{pages}{99–157} (\bibinfo{year}{1996}), ISSN \bibinfo{issn}{0550-3213}, \urlprefix\url{http://dx.doi.org/10.1016/0550-3213(95)00638-9}.

\bibitem[{\citenamefont{Kovchegov}(2000)}]{BK2}
\bibinfo{author}{\bibfnamefont{Y.~V.} \bibnamefont{Kovchegov}}, \bibinfo{journal}{Physical Review D} \textbf{\bibinfo{volume}{61}} (\bibinfo{year}{2000}), ISSN \bibinfo{issn}{1089-4918}, \urlprefix\url{http://dx.doi.org/10.1103/PhysRevD.61.074018}.

\bibitem[{\citenamefont{Kovchegov}(1999)}]{BK3}
\bibinfo{author}{\bibfnamefont{Y.~V.} \bibnamefont{Kovchegov}}, \bibinfo{journal}{Physical Review D} \textbf{\bibinfo{volume}{60}} (\bibinfo{year}{1999}), ISSN \bibinfo{issn}{1089-4918}, \urlprefix\url{http://dx.doi.org/10.1103/PhysRevD.60.034008}.

\bibitem[{\citenamefont{Hänninen et~al.}(2018)\citenamefont{Hänninen, Lappi, and Paatelainen}}]{BKd1}
\bibinfo{author}{\bibfnamefont{H.}~\bibnamefont{Hänninen}}, \bibinfo{author}{\bibfnamefont{T.}~\bibnamefont{Lappi}}, \bibnamefont{and} \bibinfo{author}{\bibfnamefont{R.}~\bibnamefont{Paatelainen}}, \bibinfo{journal}{Annals of Physics} \textbf{\bibinfo{volume}{393}}, \bibinfo{pages}{358–412} (\bibinfo{year}{2018}), ISSN \bibinfo{issn}{0003-4916}, \urlprefix\url{http://dx.doi.org/10.1016/j.aop.2018.04.015}.

\bibitem[{\citenamefont{Lappi}(2011)}]{Bkd2}
\bibinfo{author}{\bibfnamefont{T.}~\bibnamefont{Lappi}}, \emph{\bibinfo{title}{Fysh560 high energy scattering in qcd}} (\bibinfo{year}{2011}), \bibinfo{note}{lecture notes, University of Jyv{\"a}skyl{\"a}, Department of Physics}, \urlprefix\url{http://users.jyu.fi/~tulappi/fysh560kl11/hescatv2.pdf}.

\bibitem[{\citenamefont{Brodsky et~al.}(1998)\citenamefont{Brodsky, Pauli, and Pinsky}}]{Light-cone}
\bibinfo{author}{\bibfnamefont{S.~J.} \bibnamefont{Brodsky}}, \bibinfo{author}{\bibfnamefont{H.-C.} \bibnamefont{Pauli}}, \bibnamefont{and} \bibinfo{author}{\bibfnamefont{S.~S.} \bibnamefont{Pinsky}}, \bibinfo{journal}{Physics Reports} \textbf{\bibinfo{volume}{301}}, \bibinfo{pages}{299} (\bibinfo{year}{1998}), ISSN \bibinfo{issn}{0370-1573}, \urlprefix\url{https://www.sciencedirect.com/science/article/pii/S0370157397000896}.

\bibitem[{\citenamefont{Forshaw and Ross}(1997)}]{forshaw1997quantum}
\bibinfo{author}{\bibfnamefont{J.~R.} \bibnamefont{Forshaw}} \bibnamefont{and} \bibinfo{author}{\bibfnamefont{D.~A.} \bibnamefont{Ross}}, \emph{\bibinfo{title}{Quantum Chromodynamics and the Pomeron}}, Cambridge Lecture Notes in Physics (\bibinfo{publisher}{Cambridge University Press}, \bibinfo{year}{1997}), ISBN \bibinfo{isbn}{9780521568807}.

\end{thebibliography}

\end{document}